# Determination of the Hubble constant from observations of Cepheid variables in the galaxy M96

N. R. Tanvir

Institute of Astronomy, University of Cambridge, Madingley Road, Cambridge, CB3 0HA, UK.

T. Shanks

Department of Physics, University of Durham, South Road, Durham, DH1 3LE, UK.

H. C. Ferguson*

Space Telescope Science Institute, 3700 San Martin Drive, Baltimore, MD21218, USA.

D. R. T. Robinson

Institute of Astronomy, University of Cambridge, Madingley Road, Cambridge, CB3 0HA, UK.

* Hubble fellow

**New Hubble Space Telescope observations of Cepheid variable stars in the nearby galaxy M96 give a distance to the host galaxy group, Leo I, of $11.6 \pm 0.8$ Mpc. This value, used in conjunction with several reliable secondary indicators of relative distance, constrains the distances to more remote galaxy clusters, and yields a value for the Hubble constant ($H_0 = 69 \pm 8 \mathrm{kms}^{-1}\mathrm{Mpc}^{-1}$) that is independent of the velocity of the Leo I group itself.**

The value of the Hubble constant, $H_0$, has been the focus of one of the most colourful and heated debates in astrophysics. The ratio of expansion velocity to distance, $H_0$ is of fundamental importance for testing the standard Friedmann world models which have formed the essential framework of cosmology for the past half century[1,2]. Most published values fall in the range $40 < H_0 < 100 \mathrm{kms}^{-1}\mathrm{Mpc}^{-1}$, often with formal uncertainties of less than 10% (refs. 3,4,5). The intense interest in this debate stems from the fact that a value for $H_0$ greater than 75 $\mathrm{kms}^{-1}\mathrm{Mpc}^{-1}$ corresponds to an age for the universe of less than $8.7 \times 10^9$ years in the favoured Einstein–de Sitter model. This would not only be in severe conflict with the inferred ages of the oldest globular clusters[6] but would also be less than the estimated age of the galactic disk from nucleo-cosmochronology[7] and the local white dwarf cooling sequence[8].

From observations made with the Hubble Space Telescope (HST) described here, we obtain a Cepheid distance to the galaxy M96 in the Leo-I group of $11.6 \pm 0.8$Mpc. We then use the "early-type" galaxies (elliptical and lenticular) in Leo-I to deduce a distance to the Coma cluster of $105 \pm 11$ Mpc. The velocity of Coma in our microwave background rest frame is taken as $7200 \pm 300$ $\mathrm{kms}^{-1}$, where the uncertainty is intended to include allowance for the possible peculiar velocity of Coma itself[9]. Hence we find a Hubble constant of $H_0 = 69 \pm 8$ $\mathrm{kms}^{-1}\mathrm{Mpc}^{-1}$, which corresponds to an age of the universe in the standard Einstein–de Sitter ($\Omega = 1, \Lambda = 0$) cosmology of $9.5 \pm 1.1 \times 10^9$ years. Estimates of the ages of the oldest globular clusters are generally between 12 and 17 $\times 10^9$ years[6]. Only for the lowest values in this range is our result even marginally consistent with the Einstein-de Sitter model.

## HST and the distance scale

The ability of HST to resolve and monitor Cepheid variables at greater distances than previously possible represents a major opportunity to improve the local distance scale. Unfortunately, even at distances accessible to HST, galaxy "peculiar" motions are significant compared to Hubble expansion velocities[9]. It is therefore necessary to calibrate secondary distance indicators to extend the distance ladder to regimes where corrections for peculiar velocities are small.

Some secondary indicators can be calibrated directly using the spiral galaxies in which Cepheids are found. In particular, significant amounts of HST time are being devoted to calibrating the absolute magnitudes of type Ia supernovae[10] and determining the zero-point of the infrared Tully-Fisher relation[11]. However, even with these advances, long-standing disputes over the *application* of these indicators may delay consensus. The disputes centre on (1) the intrinsic scatter in the luminosity of type Ia supernovae[12,13], and the reliability of historic photometry for the calibrating supernovae[14], and (2) the intrinsic scatter of the Tully-Fisher relation[15,16].

A third and complementary route is to consider instead the various secondary indicators developed for "early-type" (elliptical and S0) galaxies. These galaxies have the important advantage over spirals in that they congregate in the cores of groups and clusters allowing many galaxies to be usefully averaged in the process of distance determination. Several of these methods have been shown to give reasonably consistent relative distances[17,18], but the lack of nearby normal giant elliptical galaxies makes calibration difficult. In the past attempts at calibration have been made using the bulges of nearby spirals, notably the Milky-Way, M31 and M81. Unfortunately there is an inevitable ambiguity when calibrating a method in a galaxy of a different type from that in which it is subsequently employed. Differences in stellar populations and the presence of disk stars and dust could introduce systematic errors.

With the HST it is now possible to pursue the alternative approach of calibrating these relations by measuring Cepheid distances to spirals in groups and clusters which also contain ellipticals. The worry here is that neighbours in projection on the sky may in fact be separated along the line of sight. This problem is illustrated by the recent work on the galaxy M100 (ref. 19). The excellent HST photometry gives a very reliable distance to M100 itself; the error in determining $H_0$ is dominated by the uncertainty in the relative position of M100 compared to the Virgo cluster core. The Virgo cluster is particularly problematic because of its known complexity of structure[20], and the evidence that the late-type members are significantly extended along the line of sight[21,22,23]. There are other, smaller, galaxy groups within reach of HST, but there is comparable uncertainty in the physical association of the spiral members[24,25,26].

The galaxy M96 (=NGC3368, $\alpha_{2000} = 10^h 46^m 45.2^s$ $\delta_{2000} = 11°49'16''$) offers the chance to break this impasse. It resides in a moderately compact group which has been variously named the M96 or Leo-I group. It is the nearest group to contain a good mix of early- and late-type galaxies (see ref 24 for a chart). The three main early-type galaxies are found to be at the same distance from the Surface Brightness Fluctuation (SBF) and Planetary Nebula Luminosity Function (PNLF) methods[18]. The group contains a unique ring of intergalactic neutral hydrogen gas[27] which, from its kinematics, is thought to be orbiting the central two early-type galaxies

M105 and NGC3384. There is a tail of gas extending from the ring towards M96 which is conventionally interpreted as evidence of an interaction[24]. M96 is already very close in both projection and velocity to the central pair, so this apparent physical connection gives us some confidence that it is at the same distance as the early-types. It is therefore an ideal location to tie together the Cepheid distance scale and the early-type galaxy scale.

## Cepheid observations

We obtained HST wide-field and planetary camera-2 (WFPC-2, ref. 28) images of M96 at 13 epochs spaced over 7 months. A 40 minute $V$-band (F555W) integration was acquired at each epoch from which we obtain the periods and average $V$ magnitudes of the variables. At three of the epochs we also obtained 50 minute $I$-band (F814W) images which are used to constrain the colours and hence account for the extinction by dust. The epochs were originally scheduled so as to sample well periods in the range 10 to 65 days. A large gap of 120 days was left between the 10th and 11th epochs to ensure that the periods for most of the variables are quite precise and free of aliases. Our strategy was altered slightly by an unscheduled HST shutdown which occurred early in our program, resulting in the rescheduling of one epoch.

The full procedure for obtaining photometry will be described in more detail elsewhere (Tanvir *et al.*, in preparation). Briefly, accurate mappings were obtained between the coordinate systems of each individual image using about 100 bright stars. A master co-added image was then created and used to obtain a coordinate list of stars down to a faint magnitude limit ($V \approx 26.2$ mag). This list was then transformed back to create a coordinate list appropriate to each frame. Aperture photometry was performed with the DAOPHOT-2 PHOT routine as implemented in IRAF[29]. Aperture corrections were determined from bright stars on the frames themselves and are accurate to $\sim 0.03$ mags[30]. We adopted the zero-points and colour terms from the WFPC-2 photometric calibration report (Holtzman *et al.* 1995, preprint). Given that uncertainties still exist in the various corrections applied to the data, we adopt an error of 0.03 mag on the zero-point calibration of each band.

Several algorithms were employed to identify variables (one discriminator is shown in figure 1). All candidate variables were tested for periodicity using a Lafler-Kinman string length statistic[31]. The Cepheid variables were selected from these on the basis of visual inspection of the light curve shape and are clearly located well above the majority of the stars in figure 1. The light curves for the Cepheids are shown in figure 2. The brightest variable in our sample probably is a Cepheid, but was observed at only 8 epochs and there is a consequent uncertainty in both its mean magnitude and period (both $\sim 80$ or $\sim 100$ days are consistent with the data). In any case, it is in the regime ($\log(P/days) > 1.8$) which was excluded in the $P$–$L$ calibration[32] since Cepheids with $\log(P) > 2$ clearly cease to lie on a linear $P$–$L$ relation. We therefore omit this variable from the sample and are left with 7 Cepheids brighter than $V = 26$ mag.

The $P$–$L$ relations for the Cepheids are shown in figure 3 where we have fixed the slope of the relations and fitted only for the zero-point[32]. This gives $\mu_{AV} = 30.51 \pm 0.12$ mag and $\mu_{AI} = 30.42 \pm 0.09$ mag, where these uncertainties come from the observed dispersion around the $P$–$L$ relations. The difference between the two apparent moduli is interpreted as evidence of reddening due to dust, corresponding to a mean total extinction of $A_V = 0.19 \pm 0.11$ mag, where the error includes the photometric calibration uncertainty. For comparison the foreground extinction to M96 from dust in the Milky-Way is estimated to be only $A_V \approx 0.05$ mag (converted from $A_B = 0.06 \pm 0.06$ mag; ref. 33). We note that although there should be a small incompleteness bias due to the magnitude limit of our current sample at $V = 26$mag, it could not be large given that our brighter Cepheids already populate the full envelope of the P–L relation as estimated from the Magellanic clouds.

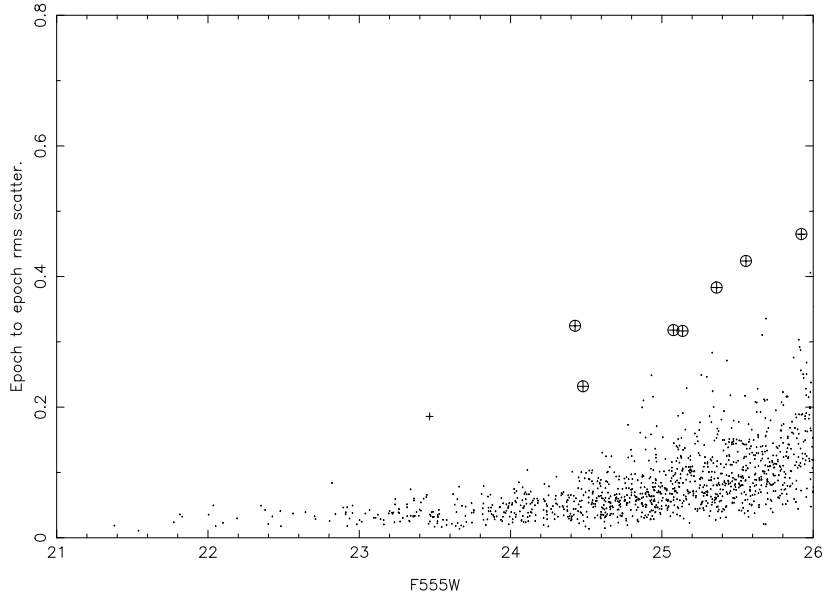

Figure 1: A plot of instrumental $V$ magnitude (filter F555W) versus the epoch to epoch $r.m.s.$ scatter. Each epoch is split into two exposures to help identify cosmic-rays, and the magnitude at each epoch is obtained in a 0.3 arcsec aperture. We selected an outer field in M96 and the level of crowding at this depth is moderate. This, combined with the poor WFPC-2 sampling, justifies the use of aperture photometry, particularly while the temporal and spatial variations of the WFPC-2 point-spread-function (psf) are not well understood[30]. The median $r.m.s.$ scatter at $V = 26$mag is about 0.18mag, similar to that obtained in M100 (ref. 19), although this varies somewhat with position on the frame due to the changing background level across the field. The eight Cepheids are marked and are clearly separated from the main locus of stars. The brightest Cepheid, marked with an uncircled cross, was observed at only 8 epochs and its period and magnitude are correspondingly uncertain. Some of the other stars with high scatter may be other types of variables, although a few outliers are expected from noise and occasional contamination of stars by "hot pixels" or cosmic rays. The Cepheids can also be identified by the shape of their light curves.

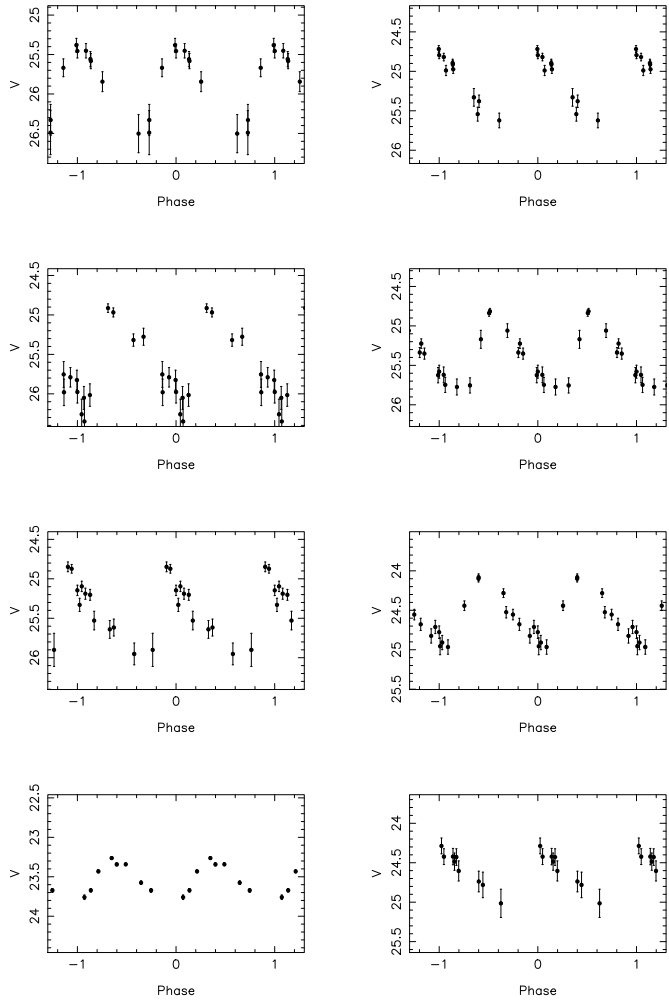

Figure 2: Light curves of the eight Cepheids folded on the best periods. The amplitudes of variation are comparable with those seen in studies of nearer galaxies (eg. ref 11) indicating that there is unlikely to be a problem with field stars overlapping the Cepheid images. Approximate periods were determined by a number of automated methods, all of which agreed well with each other. Ultimately the visual fit was considered most reliable although in no case did this make a significant difference. The well-known "saw-tooth" signature of classical Cepheids is evident. The telescope pointing was shifted by a small amount between most epochs and, in particular, the final three epochs were obtained at a position angle approximately 180° from the others. Thus some stars were not imaged at all epochs. In addition, the very large number of cosmic rays and hot pixels on many of the images meant that, although each epoch was split into two exposures, occasionally a star was "hit" in both, and hence no magnitude found for that epoch. All the Cepheids used in fitting the $P$–$L$ relations had good photometry for at least 9 epochs. Only for the longest period variable (bottom left plot) is the sampling too poor to be sure of the magnitude or period (there is an alias at $\sim 100$ days).

|     | Source of uncertainty | R.M.S. magnitude error |
|-----|------------------------|------------------------|
| [A] | Adopted V & I zero-point error | 0.03 |
| [B] | Spread of reddening corrected distance moduli from 7 Cepheids | 0.21 |
| [C] | Uncertainty in LMC to Leo modulus ($2.45 \times$[A], $1.45 \times$[A], [B]/$\sqrt{7}$) | 0.12 |
| [D] | Uncertainty in LMC distance modulus[32] (ie. the Cepheid calibration) | 0.10 |
| [E] | Allowance for possible metallicity effect on Cepheid P–L relation | 0.05 |
| [F] | Uncertainty in M96 distance modulus ([C], [D], [E]) | 0.16 |
| [G] | Estimated r.m.s. depth of Leo-I group (see text) | 0.04 |
| [H] | Uncertainty in Leo-I group centroid ([F], [G]) | 0.17 |
| [I] | Adopted Leo-I to Virgo uncertainty from figure 4 | 0.15 |
| [J] | Adopted Virgo to Coma uncertainty (see text) | 0.10 |
| [K] | Leo-I to Coma - route via Virgo cluster ([I], [J]) | 0.18 |
| [L] | Leo-I to Coma - route via $D_n - \sigma$ relation | 0.34 |
| [M] | Combined Leo to Coma from weighted mean of [K] and [L] | 0.16 |
| [N] | Resultant error in Coma modulus ([H], [M]) | 0.23 |

Table 1: This table shows a complete error budget used in deriving the distance to Coma. Figures in parentheses are added in quadrature to obtain the result. Numbers are generally rounded to two significant figures. Note that our step [B], in which we find the spread in the 7 independent estimates of $\mu_0 = 2.45 \times \mu_{AI} - 1.45 \times \mu_{AV}$, automatically accounts for the correlation between the V and I residuals from the P–L relations. The main difference between our budget and that of Freedman et al.[19] for M100, is in the allowance made for the depth of the group. Clearly the strength of our result for $H_0$ depends on the validity of the arguments made for M96 being closely associated with the Leo-I ellipticals. Note also that we are using the most recent WFPC-2 calibration by Holtzman et al. (preprint), which is more precise than the provisional calibration available when the M100 result was published. This largely accounts for our improved precision in obtaining the distance to M96 itself. The reddening relation $A_V = 2.45 E_{V-I}$ also comes from Holtzman et al.

The calibration of the period-luminosity relations, used above, is based on observations of Cepheids in the Magellanic clouds[32] and assumes an LMC distance modulus of $\mu_0 = 18.5 \pm 0.1$ mag and reddening $E(B-V) = 0.1$ mag. Combining the total internal error with the uncertainties in the photometric calibration gives $\mu_0(\text{Leo}) - \mu_0(\text{LMC}) = 11.82 \pm 0.12$ mag. Hence we obtain a true distance modulus for M96 itself of $\mu_0 = 30.32 \pm 0.16$ mag, which corresponds to a distance of $11.6 \pm 0.8$ Mpc. Any effect of differences in metal abundance between the LMC and our M96 field is expected to be small for $V - I$ colours[34], and we have included an uncertainty of 0.05 mag to allow for this.

The Leo-I group is $\sim 6°$ across, although most "high-confidence" members are in a $2.5° \times 1.5°$ region. If we assume rough spherical symmetry, this would imply an r.m.s. depth of $\sim 2\%$ of the distance to the group. In particular M96, given its apparent interaction with the HI ring surrounding the central galaxies, is probably even closer to the centroid. Therefore the depth of the group is likely to be negligible compared to the uncertainty in the distance. A summary of the error budget for our $H_0$ estimate is given in table 1.

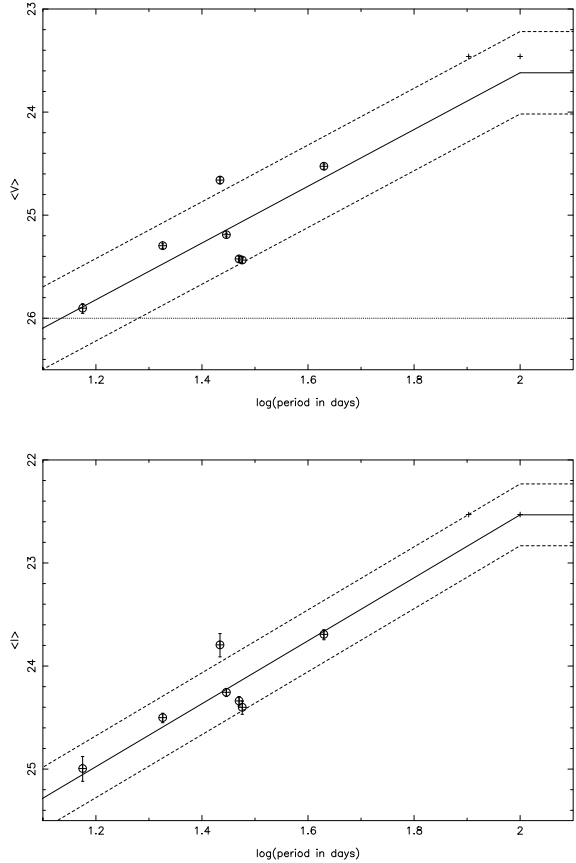

Figure 3: Here we fit the Cepheid period-luminosity relations of Madore and Freedman[32] to both the $V$ and $I$ data. For some time the calibrations of the Cepheid variable period–luminosity ($P$–$L$) relations have been in good agreement, as shown by the various Galactic calibrations[39], and the several independent distances to the Magellanic Clouds, which are generally consistent with their Cepheid distances[4,40]. The level of agreement is illustrated by the common use of the Madore and Freedman[32] $P$–$L$ relations by advocates of both low[10] and high[38] values of $H_0$. The average Cepheid magnitudes are defined as the magnitude at mean intensity. The slopes are fixed to be 2.76 in $V$ and 3.06 in $I$ from the Magellanic cloud Cepheids[32] so there is only one free parameter in each case. The approximate envelopes of the Magellanic cloud Cepheids are also plotted. Only seven Cepheids are included in the fit since the brightest variable is of uncertain period (it is plotted with both 80 and 100 days) and is close to the point where the $P$–$L$ relations become non-linear and not well understood. The error bars are derived only from photon counting statistics and therefore do not account for errors in fitting the light curves and errors due to crowding, however they provide reasonable weights for the $P$–$L$ fits. The effective magnitude limit at $V = 26$ is marked; we expect any bias due to incompleteness to be small.

## From Leo-I to the Coma cluster

The Leo-I group is relatively nearby and so its Hubble recession velocity is made uncertain by local peculiar motions[9]. We therefore consider two routes to step from the Leo-I group to the more distant Coma cluster, whose peculiar velocity is small compared to its Hubble velocity (figure 4). Firstly, we use the Virgo cluster as a stepping stone by adopting the Leo-I to Virgo distance ratio obtained from the weighted mean of a variety of distance indicators. These are summarized in figure 4 and give $\mu_0(\text{Virgo}) - \mu_0(\text{Leo}) = 0.99 \pm 0.15$ mag. The indicators with the highest claimed precision, and hence greatest weight, are the $I$-band SBF and the PNLF methods. The disagreement between these two methods, although small, appears to be significant given the quoted errors. There are reasons to suspect that the problem could lie with the PNLF method which is reaching its limit at Virgo cluster distances and may also have a dependence on parent galaxy luminosity[18,35]. However, our adopted uncertainty of 0.15 mag, which is more than three times the formal error, encompasses both results at the $1\sigma$ level. Although we are determining $H_0$ at the Coma cluster, we note in passing that the above implies a Virgo distance of $18.3 \pm 2.0$Mpc which is intermediate between the traditional "short-scale" of 13–16Mpc and "long-scale" of 20–24Mpc.

The published estimates for the relative Virgo–Coma distance are generally in good agreement. Rather than summarizing all the available measures of this ratio the reader is referred to recent reviews[3-5,36], which conclude values for $\mu_0(\text{Coma}) - \mu_0(\text{Virgo})$ of 3.80, 3.71, 3.69–3.74 and 3.80 mag respectively. We therefore adopt $3.75 \pm 0.1$ mag with, again, a reasonably conservative uncertainty. Combining this with the Leo-Virgo relative distance gives $\mu_0(\text{Coma}) - \mu_0(\text{Leo}) = 4.74 \pm 0.18$ mag.

A second route to Coma is to establish the zero-point of the $D_n - \sigma$ relation[37] (a relation between the diameter and the internal velocity dispersion for elliptical galaxies) in the Leo-I group itself using the two ellipticals M105 and NGC3377. This has a higher uncertainty due to the intrinsic scatter around the $D_n - \sigma$ relation, but offers a check which does not make use of the somewhat newer SBF and PNLF methods. The relative distance between these two Leo-I galaxies and the mean for the Coma cluster is $\mu_0(\text{Coma}) - \mu_0(\text{Leo}) = 4.90 \pm 0.34$ (ref. 37). While the $D_n - \sigma$ relation is also amongst the techniques used to obtain the Virgo to Coma distance ratio, because the uncertainty is dominated by the small number of Leo-I galaxies this route to Coma is largely independent of the route via Virgo. Clearly the two routes agree well within their respective errors. Taking a weighted average of these results and our Cepheid distance to M96 we obtain a distance modulus to Coma of $35.10 \pm 0.23$ mag or a distance of $105 \pm 11$ Mpc. Hence we obtain $H_0 = 69 \pm 8$ kms$^{-1}$Mpc$^{-1}$as described above.

## Future prospects

Our value of the Hubble constant is based on the standard Cepheid *P–L* relations and the reasonable assumption of small peculiar velocity corrections at Coma. The intermediate step relies on calibrating the early-type galaxy distance scale in Leo-I and thus avoids some of the disputes which have contributed to the long-standing distance scale controversy. The result still has an uncertainty which is consistent at the $2\sigma$ level with either the "high" values of Pierce *et al.*[38] and Freedman *et al.*[19], or the "low" value of Sandage *et al.*[10]. The size of the uncertainty depends on the arguments for M96 being in close proximity to the Leo-I early-type galaxies. Ongoing HST observations of UGC5889 and M95 (by our group and the key-project team) should help to test whether there is any significant extension along the line of sight and to refine the mean distance to the group.

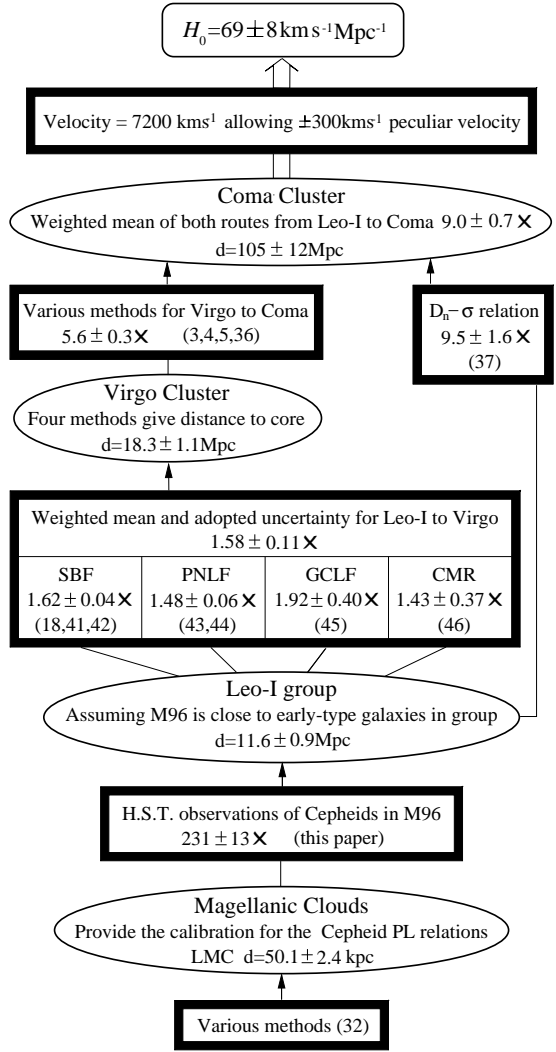

Figure 4: This figure shows all the steps involved in deriving $H_0$. Each box represents a distance ratio as inferred from data given in the papers referenced in parentheses. We start with the standard Cepheid $P$–$L$ relations obtained in the Magellanic clouds[32] and step via Leo-I to the Coma cluster. The Coma cluster is assumed to be sufficiently remote that the corrections to its velocity for peculiar motions are small[9]. The two routes from Leo-I to Coma rely primarily on various early-type galaxy distance indicators, namely the surface brightness fluctuation (SBF), planetary nebula luminosity function (PNLF), globular cluster luminosity function (GCLF), colour-magnitude relation (CMR) and $D_n - \sigma$ methods. Note that the $D_n - \sigma$ relation gives $d(\text{Virgo})/d(\text{Leo}) = 1.71 \pm 0.30$ (ref. 37) but is omitted in the route via Virgo so that the direct Leo-I to Coma route is kept as independent as possible. Although the formal error on the weighted mean in the Leo-I to Virgo step is only $\approx 2\%$ we have allowed a larger uncertainty to account for possible systematic effects. As the early-type indicators are fairly consistent with each other, the value of $H_0$ we derive is reasonably insensitive to the choice of weighting.

Otherwise the error estimate makes reasonable allowance for many possible sources of external error and is not dominated by any single uncertainty. Further measurements with the various early-type galaxy distance indicators are required to improve the relative Leo-I to Coma distances and also to establish whether there is a significant discrepancy between the SBF and PNLF methods. With these improvements, HST distances to Leo-I group galaxies offer the prospect of a new and relatively secure route to determining $H_0$.

Acknowledgements.

Particular thanks go to Mike Irwin for obtaining a ground-based image of M96 which was useful in selecting the HST field centre; to the staff at STScI, especially Doug van Orsow, for assistance in planning and scheduling the HST observations; to Jon Holtzman for clarification of WFPC-2 calibration; to Donald Lynden-Bell for useful discussions; and Mario Livio for a critical reading of the manuscript. This work was based on observations with the NASA/ESA Hubble Space Telescope, obtained at the Space Telescope Science Institute which is operated by the Association of Universities for Research in Astronomy.